\documentclass[aps,prd,twocolumn,amsmath,superscriptaddress,amssymb,showpacs,floatfix,nofootinbib,longbibliography]{revtex4-1}
\pdfoutput=1
\usepackage{graphicx}
\usepackage{bm}
\usepackage{times}
\usepackage{hyperref}
\usepackage{slashed}
\usepackage{color}
\usepackage{aas_macros}
\usepackage{slashed}
\usepackage{lipsum}
\usepackage{subfigure}
\usepackage{multirow}
\usepackage{amsmath}
\usepackage{array} 
\usepackage{varwidth} 
\usepackage{enumitem}

\usepackage{scalerel}
\usepackage{tikz}
\usetikzlibrary{svg.path}

\definecolor{orcidlogocol}{HTML}{A6CE39}
\tikzset{
	orcidlogo/.pic={
		\fill[orcidlogocol] svg{M256,128c0,70.7-57.3,128-128,128C57.3,256,0,198.7,0,128C0,57.3,57.3,0,128,0C198.7,0,256,57.3,256,128z};
		\fill[white] svg{M86.3,186.2H70.9V79.1h15.4v48.4V186.2z}
		svg{M108.9,79.1h41.6c39.6,0,57,28.3,57,53.6c0,27.5-21.5,53.6-56.8,53.6h-41.8V79.1z M124.3,172.4h24.5c34.9,0,42.9-26.5,42.9-39.7c0-21.5-13.7-39.7-43.7-39.7h-23.7V172.4z}
		svg{M88.7,56.8c0,5.5-4.5,10.1-10.1,10.1c-5.6,0-10.1-4.6-10.1-10.1c0-5.6,4.5-10.1,10.1-10.1C84.2,46.7,88.7,51.3,88.7,56.8z};
	}
}

\newcommand\orcidicon[1]{\href{https://orcid.org/#1}{\mbox{\scalerel*{
				\begin{tikzpicture}[yscale=-1,transform shape]
				\pic{orcidlogo};
				\end{tikzpicture}
			}{1}}}}

\newcommand{\be}{\begin{equation}}
\newcommand{\ee}{\end{equation}}
\newcommand{\bea}{\begin{eqnarray}}
\newcommand{\eea}{\end{eqnarray}}

\let\vec\bm
\newcommand{\uvec}[1]{\bm{\hat{#1}}}

\newcommand{\diff}{\ensuremath{\mathrm{d}}}
\newcommand{\TRH}{\ensuremath{T_\mathrm{RH}}}

\newcommand{\kcut}{\ensuremath{k_\mathrm{cut}}}
\newcommand{\kRH}{\ensuremath{k_\mathrm{RH}}}

\newcommand{\xcut}{\ensuremath{\kcut/\kRH}}
\newcommand{\rmax}{\ensuremath{r_\mathrm{max}}}
\newcommand{\mmax}{\ensuremath{M_\mathrm{max}}}
\newcommand{\e}{\mathrm{e}}
\newcommand{\Msun}{\ensuremath{\mathrm{M}_\odot}}
\newcommand{\tobs}{\ensuremath{t_\mathrm{obs}}}
\newcommand{\tres}{\ensuremath{t_\mathrm{res}}}
\newcommand{\tcad}{\ensuremath{t_\mathrm{cad}}}

\begin{document}

\title{Dark Matter Microhalos in the Solar Neighborhood: \\ Pulsar Timing Signatures of Early Matter Domination}
\author{M. Sten Delos \orcidicon{0000-0003-3808-5321}}
\email{sten@mpa-garching.mpg.de}
\affiliation{Max Planck Institute for Astrophysics, Karl-Schwarzschild-Stra{\ss}e 1, 85748 Garching, Germany}
\author{Tim Linden \orcidicon{0000-0001-9888-0971}}
\email{linden@fysik.su.se}
\affiliation{Stockholm University and The Oskar Klein Centre for Cosmoparticle Physics,  Alba Nova, 10691 Stockholm, Sweden}

\begin{abstract}
Pulsar timing provides a sensitive probe of small-scale structure. Gravitational perturbations arising from an inhomogeneous environment could manifest as detectable perturbations in the pulsation phase. Consequently, pulsar timing arrays have been proposed as a probe of dark matter substructure on mass scales as small as $10^{-11}~\Msun$. Since the small-scale mass distribution is connected to early-Universe physics, pulsar timing can therefore constrain the thermal history prior to Big Bang nucleosynthesis (BBN), a period that remains largely unprobed. We explore here the prospects for pulsar timing arrays to detect the dark substructure imprinted by a period of early matter domination (EMD) prior to BBN. EMD amplifies density variations, leading to a population of highly dense sub-Earth-mass dark matter microhalos. We use recently developed semianalytic models to characterize the distribution of EMD-induced microhalos, and we evaluate the extent to which the pulsar timing distortions caused by these microhalos can be detected. Broadly, we find that sub-0.1-$\mu$s timing noise residuals are necessary to probe EMD. However, with 10-ns residuals, a pulsar timing array with just 70 pulsars could detect the evidence of an EMD epoch with 20~years of observation time if the reheat temperature is of order 10~MeV. With 40 years of observation time, pulsar timing arrays could probe EMD reheat temperatures as high as 150~MeV.
\end{abstract}

\maketitle

\section{Introduction}

Pulsar timing presents an exquisitely sensitive probe of gravitational perturbations. Due to the highly regular periodicity of many pulsars' emissions, tiny disturbances can manifest as detectable shifts in their pulsation phases. For this reason, pulsar timing is indispensable in the search for low-frequency gravitational waves, for which pulsar timing arrays (PTAs)---which search for correlated phase shifts---represent the most sensitive detection methodology~\cite{Verbiest:2021kmt}. However, PTAs can also probe perturbations arising from an inhomogeneous matter environment. In particular, pulsar timing holds the potential to detect primordial black holes~\cite{Seto:2007kj, 2012MNRAS.426.1369K, Schutz:2016khr} and dark matter subhalos~\cite{Siegel:2007fz, Baghram:2011is, clark2015investigatingI,*clark2016erratumI,2018arXiv180107847K, Dror_2019,Ramani_2020,lee2021probing,2021arXiv210405717L} as small as $10^{-11}~\Msun$, which lie far beyond the reach of other astrophysical probes of dark matter.

The small-scale distribution of cold dark matter is closely linked to the Universe's early history. The cosmic microwave background (CMB) and light element abundances establish that the Universe was dominated by the Standard Model radiation bath by the time it cooled to a temperature of a few MeV~\cite{2019JCAP...12..012H,2015PhRvD..92l3534D}. However, there is little reason to assume that radiation domination extends to higher temperatures (i.e., earlier times); see Ref.~\cite{2021OJAp....4E...1A} for a review. In particular, a well motivated possibility is that the Universe was dominated at earlier times by an unstable heavy field. Examples of species that could drive this period of early matter domination (EMD) include hidden-sector particles \cite{Pospelov_2008,Arkani_Hamed_2009,Hooper_2012,Abdullah_2014,Berlin_2014,Martin_2014,zhang2015long,Berlin_2016a,Berlin_2016b,Dror_2016,Tenkanen:2016jic,Dror_2018,Tenkanen:2019cik}, moduli fields in string theory \cite{coughlan1983cosmological,de1993model,banks1994cosmological,banks1995cosmological,banks1995modular,acharya2014bounds,Kane_2015,Giblin_2017}, inflationary spectator fields \cite{mollerach1990isocurvature,linde1997non,lyth2002generating,moroi2001effects,*moroi2002erratum}, and the inflaton itself \cite{albrecht1982reheating,turner1983coherent,traschen1990particle,kofman1994reheating,kofman1997towards,dufaux2006preheating,allahverdi2010reheating,jedamzik2010collapse,easther2011delayed,musoke2019lighting}. EMD has a marked impact on the small-scale mass distribution because the gravitational clustering of the early matter particle boosts density variations on scales that were subhorizon (i.e., causal) during this time \cite{erickcek2011reheating}. These density variations could persist within the dark matter distribution even after the early matter species decays.

In this article, we explore the potential for pulsar timing to probe an early matter-dominated epoch through its imprint on small-scale dark structure. The boost to density variations causes a large fraction of the dark matter to become bound into highly dense sub-earth-mass microhalos \cite{erickcek2011reheating}. If a microhalo passes near a pulsar (or the Earth), it perturbs the body's motion and potentially leads to a detectable pulsar timing distortion. Sufficiently compact microhalos that cross the line of sight to a pulsar can also perturb the incoming light directly. Recently, Ref.~\cite{lee2021probing} studied microhalo detection prospects using pulsar timing arrays; they considered EMD-induced microhalos (among others) and concluded that an EMD could be probed if it ends at a reheat temperature $\TRH$ below about 1~GeV. Our research builds on the analysis of Ref.~\cite{lee2021probing}. In particular, microhalos detectable by pulsar timing are near the Galactic disk, making them susceptible to disruption by tidal forces and high-speed encounters. Within this context, we employ recently developed semianalytic descriptions of microhalo formation and evolution to precisely characterize the EMD-induced microhalo distribution that manifests within this environment. We also explore the parameter space for early matter-dominated epochs more exhaustively.

To describe microhalo formation, we employ the approach developed in Ref.~\cite{delos2019predicting}. This prescription maps peaks in the initial density field onto collapsed halos at later times; the properties of the peak predict the internal structure of the resulting halo. The next step is to model how microhalos respond to the Galactic disk's disruptive environment, and for this purpose we employ the model presented in Ref.~\cite{delos2019tidal} to describe subhalo tidal evolution and that presented in Ref.~\cite{delos2019evolution} to describe the impact of high-speed encounters. We previously used these models in Ref.~\cite{delos2019breaking} to characterize another potential signature of EMD---boosted dark matter annihilation---and we also employ here the model refinements developed for that work. \textsc{Python} codes that implement the microhalo models employed in this work are publicly available.\footnote{\url{https://github.com/delos/microhalo-models}}

In this paper, we parametrize the impact of an arbitrary EMD scenario (following Ref.~\cite{2015PhRvD..92j3505E}) in terms of its reheat temperature $\TRH$ and the ratio $\xcut$ between the dark matter free-streaming wavenumber and the wavenumber that enters the horizon at reheating. Both of these parameters are necessary to characterize the imprint of early matter domination: the former sets the scales at which density variations are boosted, while the latter sets the amplitude of the boost. On the observational side we consider the impact of the pulsar count $N_P$ and observation duration $\tobs$ as well as the rms timing noise residual $\tres$. We find that, roughly, timing residuals $\tres\lesssim 0.1~\mu$s are needed to detect EMD epochs. If $\tres=10$~ns, EMD-induced microhalos could be detected with about $N_P=70$ pulsars and $\tobs\simeq 20$~years if $\TRH\lesssim 20$~MeV and $\xcut\gtrsim 30$. Higher $\TRH$ and lower $\xcut$ can be reached with larger $\tobs$ and $N_P$, although microhalo detection prospects are relatively insensitive to $N_P$. With hundreds to thousands of pulsars, reheat temperatures as high as $\TRH\simeq 150$~MeV or cutoff ratios as low as $\xcut\simeq 8$ could be probed in $\tobs=40$~years.

This article is organized as follows. Section~\ref{sec:power} describes the matter power spectrum that arises from an EMD scenario. In Sec.~\ref{sec:initial}, we characterize the microhalo distribution resulting from such a power spectrum. In Sec.~\ref{sec:disruption}, we treat the evolution of microhalos due to tidal forces and encounters within the Galactic potential. Section~\ref{sec:procedure} outlines our method for assessing the detection prospects of a given microhalo population using pulsar timing, which is largely the same as the method presented in Ref.~\cite{lee2021probing}. These detection prospects, which constitute the main results of this article, are presented in Sec.~\ref{sec:results}. Section~\ref{sec:conclusion} presents our conclusions. Finally, Appendix~\ref{sec:Vrel} further details Sec.~\ref{sec:disruption}'s computation of the microhalo-stellar encounter distribution, while Appendix~\ref{sec:profile} refines the model in Ref.~\cite{delos2019evolution} to better describe a halo's long-term response to a high-speed encounter.

\section{Density variations arising from EMD}
\label{sec:power}

We first review the spectrum of density variations that results from an early matter-dominated phase. During EMD, the dominant species, which we call $\phi$, gravitationally clusters and creates gravitational potential wells. If the dark matter is both nonrelativistic and kinetically decoupled from any relativistic species (such as the Standard Model) by a temperature of about $2\TRH$ \cite{delos2019breaking}, then it falls into these potential wells and inherits the density variations in $\phi$. After reheating, these density variations persist in the dark matter. 

\begin{figure}[tbp]
	\centering
	\includegraphics[width=\columnwidth]{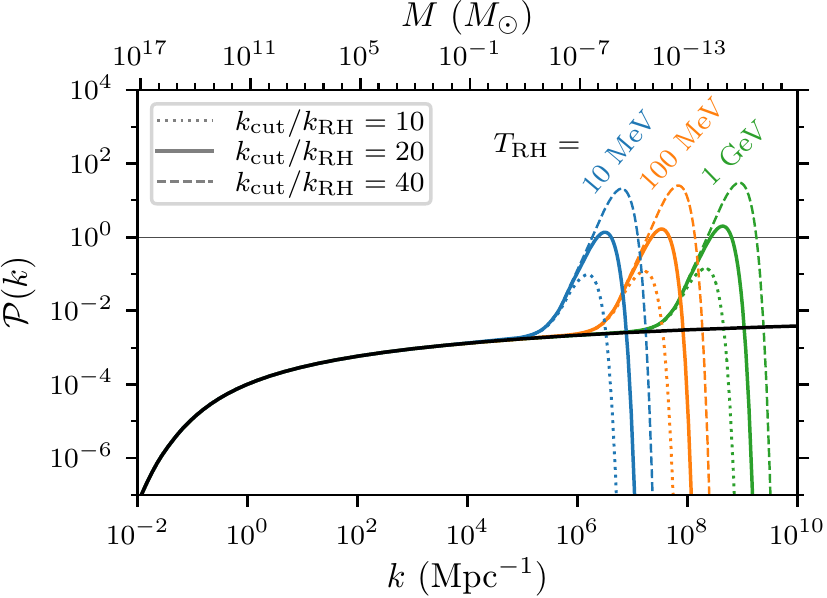}
	\caption{EMD-induced (dimensionless) dark matter power spectra at $z=300$, computed using linear theory. We show different values of the reheat temperature $\TRH$ and the dark matter free-streaming cutoff scale $\kcut$, expressed as the ratio $\xcut$. $\TRH$ sets the scales at which EMD boosts density variations, while $\xcut$ sets the maximum amplitude of the boost. For comparison, the solid black curve indicates the cold dark matter power spectrum without early matter domination. We also show the mass scale $M$ associated with each wavenumber, which roughly corresponds to the masses of halos forming from density variations of that scale. For $\xcut\gtrsim 20$, density variations are already nonlinear ($\mathcal{P}(k)\gtrsim 1$), implying microhalos are already forming by $z=300$.}
	\label{fig:pk}
\end{figure}

Reference~\cite{erickcek2011reheating} developed a prescription to quantify the power spectrum $\mathcal{P}(k)$ of density fluctuations imprinted onto the dark matter by a general EMD epoch. The form of $\mathcal{P}(k)$ is influenced by two parameters: the reheat temperature $\TRH$ associated with the end of EMD; and the dark matter free-streaming scale, which sets a cutoff wavenumber $\kcut$.\footnote{The $\TRH$--$\kcut$ parametrization is mostly general; it only neglects the possibility of a very short EMD epoch preceded by a radiation-dominated (or other) epoch. In particular, we assume that $\kcut$ is smaller than the wavenumber that enters the horizon at the beginning of EMD. Otherwise, a third parameter would be necessary to represent that wavenumber.} Figure~\ref{fig:pk} shows the dark matter power spectrum predicted by this prescription for several different EMD scenarios. These power spectra are computed in linear theory, i.e., assuming $\mathcal{P}(k)\ll 1$. A key observation is that if the ratio $\xcut$ is larger than about 20 between $\kcut$ and the wavenumber, $\kRH$, which enters the horizon at reheating, then density variations are already becoming nonlinear ($\mathcal{P}\gtrsim 1$) by the redshift $z=300$ at which the power spectra are plotted. Consequently, collapsed halos are already beginning to form at this early time. We also emphasize the fundamental importance of the cutoff scale $\kcut$ to the EMD epoch's imprint on density variations: if $\kcut<\kRH$, then free streaming erases any trace of EMD from the dark matter distribution.

We focus in this study on the reheat-temperature range $3~\text{MeV}<\TRH<180~\text{MeV}$. According to Ref.~\cite{2015PhRvD..92l3534D}, the effective number of neutrino species reflected in the CMB constrains $\TRH>4.7$~MeV, which motivates our lower bound. Meanwhile, high reheat temperatures $\TRH$ are associated with smaller mass scales (see Fig.~\ref{fig:pk}), which are more difficult to detect through pulsar timing, and we will see later that reheat temperatures $\TRH\gtrsim 180$~MeV are broadly inaccessible. With respect to the cutoff parameter $\xcut$,\footnote{Note that the two scales $\kRH$ and $\kcut$ are naturally connected despite being associated with entirely different physics ($\kRH$ with the $\phi$ decay rate and $\kcut$ with dark matter's interaction properties). To achieve the observed dark matter abundance, a thermal relic must thermally decouple from the radiation bath not too long before reheating; see Ref.~\cite{delos2019breaking}. The free-streaming scale $\kcut$ is set by the the time of \textit{kinetic} decoupling, which is in turn related to the time of \textit{thermal} decoupling through the particular dark matter microphysics. Further details of the $\kcut$--$\kRH$ connection are laid out in Refs.~\cite{2015PhRvD..92j3505E,2016PhRvD..94f3502E}, the latter of which suggests that $\xcut$ as high as 200 can be plausible for supersymmetric dark matter. However, the presence of a radiation-dominated epoch prior to EMD can break the $\kcut$--$\kRH$ connection and allow $\kcut\gg\kRH$ (see, e.g., Ref.~\cite{2019PhRvD.100j3010B}).} we limit our consideration to $5<\xcut<40$. We will see that microhalos arising in scenarios with $\xcut<5$ are not dense enough to be detected. On the other hand, larger values $\xcut>40$ lead to the collapse of overdense regions (and likely halo formation \cite{2019PhRvD.100j3010B}) deep in the radiation-dominated epoch, a scenario for which our modeling of halo populations is not calibrated. Nevertheless, as we discuss in Sec.~\ref{sec:results}, our conclusions are likely applicable to many scenarios with $\xcut>40$.

\section{The initial microhalo distribution}\label{sec:initial}

To understand the present-day imprint of EMD, the next step is to advance from the linear-theory power spectrum to the population of (nonlinear) collapsed dark matter halos. To model halo populations in this way, it is common to use Press-Schechter theory \cite{press1974formation,bond1991excursion}. In this model, a halo of mass $M$ is associated with a region of Lagrangian space (the space of initial comoving particle positions) of mass $M$ inside which the average linear-theory density contrast $\delta(\vec x)\equiv \delta \rho/\rho$ exceeds some critical value of order 1. Mathematically, the linear-theory density field $\delta(\vec x)$ is convolved with a filter that smooths it over the mass scale $M$.

However, the smallest dark matter microhalos form directly from the collapse of peaks in the linear-theory density field $\delta(\vec x)$. To study them, it is more natural to consider the unfiltered density field and simply map each peak therein to a collapsed halo. This procedure carries the added benefit that each halo's density profile can be predicted from the properties of the associated peak \cite{delos2019predicting}, so we do not need to rely on empirical concentration-mass relations (e.g., Ref.~\cite{ludlow2016mass}) that can be difficult to generalize to arbitrary power spectra.

Therefore we choose to characterize the microhalo population using the peak-based prescription that Ref.~\cite{delos2019predicting} introduced.\footnote{This prescription was also employed in the context of EMD by Ref.~\cite{delos2019breaking} to study a different potential observable signature---enhanced annihilation radiation.} To begin, we exploit the statistics of Gaussian fields \cite{bardeen1986statistics} to generate a random sample of peaks along with their density profiles $\delta(q)$, where $q$ is the comoving radius. With these peak density profiles, we estimate the density profiles of the resulting collapsed halos using a secondary infall model \cite{gunn1972infall,gott1975formation}. Specifically, we use the simulation-tuned ``turnaround'' model of Ref.~\cite{delos2019predicting} (refined as described in Appendix~A of Ref.~\cite{delos2019breaking}) to predict the radius $\rmax$ at which the circular velocity is maximized and the corresponding enclosed mass $\mmax$. Some peaks do not have a predicted collapse time due to their high ellipticity, and we neglect these peaks.\footnote{Collapse time is predicted using the approximation in Ref.~\cite{sheth2001ellipsoidal}, which is not guaranteed to have a solution. Whether these peaks actually fail to collapse is likely immaterial, as ellipticity is anticorrelated with peak height, so the alternative is that these peaks collapse late and produce halos of low density.} Additionally, some peaks produce halos whose predicted $\rmax$ values are so large as to be populated by material initially lying well beyond the scales boosted by early matter domination. These predictions represent conventional CDM halos unboosted by EMD, so they are irrelevant to pulsar timing observations (e.g., Ref.~\cite{lee2021probing}), and we neglect them for computational convenience.\footnote{The $\rmax$ and $\mmax$ predictions involve the covariance matrix between the field values at all different radii about the peak, so it is computationally beneficial to limit the radial extent explored.}

While microhalos form with density profiles that asymptote to $\rho\propto r^{-3/2}$ at small radii \cite{ishiyama2010gamma,anderhalden2013density,*anderhalden2013erratum,ishiyama2014hierarchical,polisensky2015fingerprints,ogiya2017sets,delos2018ultracompact,delos2018density,angulo2017earth,delos2019predicting,ishiyama2019abundance}, successive mergers drive their inner density cusps toward a shallower $\rho\propto r^{-1}$ scaling \cite{ogiya2016dynamical,angulo2017earth,gosenca20173d,delos2019predicting,ishiyama2019abundance}. Therefore, we assume microhalos develop the Navarro-Frenk-White (NFW) profile \cite{navarro1996structure,navarro1997universal},
\be\label{NFW}
\rho(r) = \frac{\rho_s}{(r/r_s)(1+r/r_s)^{2}},
\ee
where the scale parameters $r_s$ and $\rho_s$ are set to reproduce the predicted $\rmax$ and $\mmax$. No precise description of the impact of microhalo mergers has yet been presented; the above procedure yields a conservative estimate, as mergers between microhalos are understood to raise $\rmax$ and $\mmax$ \cite{delos2019predicting}.

The above prescription sets the distribution of microhalo density profiles. We set their total number density in the following way. The cosmological microhalo number density $\bar n$ is simply the number density of initial density peaks (which can be computed from $\mathcal{P}(k)$ using, e.g., the methods of Ref.~\cite{bardeen1986statistics}) scaled to account for the aforementioned removal of some peaks from consideration. The number density of microhalos in the Sun's vicinity is then $n = (\rho_\mathrm{local}/\bar\rho)\bar n$, where $(\rho_\mathrm{local}/\bar\rho)\simeq 3\times 10^5$ is the factor by which the local dark matter density exceeds the cosmological mean.

These procedures do not fully account for the impact of microhalo-microhalo mergers, a topic that we leave for future study. These mergers tend to raise microhalo masses, as noted above, while also reducing their number count. While these two effects have opposite implications on observational prospects, we argue that neglecting both of them represents a conservative choice. Mergers have two relevant effects:
\begin{enumerate}[label={(\arabic*)}]
	\item Mergers boost individual microhalo masses while preserving their overall mass content (i.e., the fraction of total dark matter mass residing in microhalos).
	\item Mergers tend to approximately preserve halos' internal density values (i.e., concentrations) \cite{delos2019predicting,drakos2019major}.
\end{enumerate}
Reference~\cite{lee2021probing}, exploring constraints from pulsar timing on monochromatic halo mass functions, found that when condition (2) holds---concentrations are held fixed---the parameter change (1)---increasing the halo mass while preserving the fraction of total dark matter mass residing in microhalos---can only strengthen pulsar timing constraints. Consequently, mergers between microhalos can only boost their detectability in pulsar timing. We also note that the preservation of microhalo internal densities implies that their susceptibility to disruptive effects within the Galaxy, which we treat in the next section, is unaltered by mergers.

\begin{figure*}[tbp]
	\centering
	\includegraphics[width=.8\linewidth]{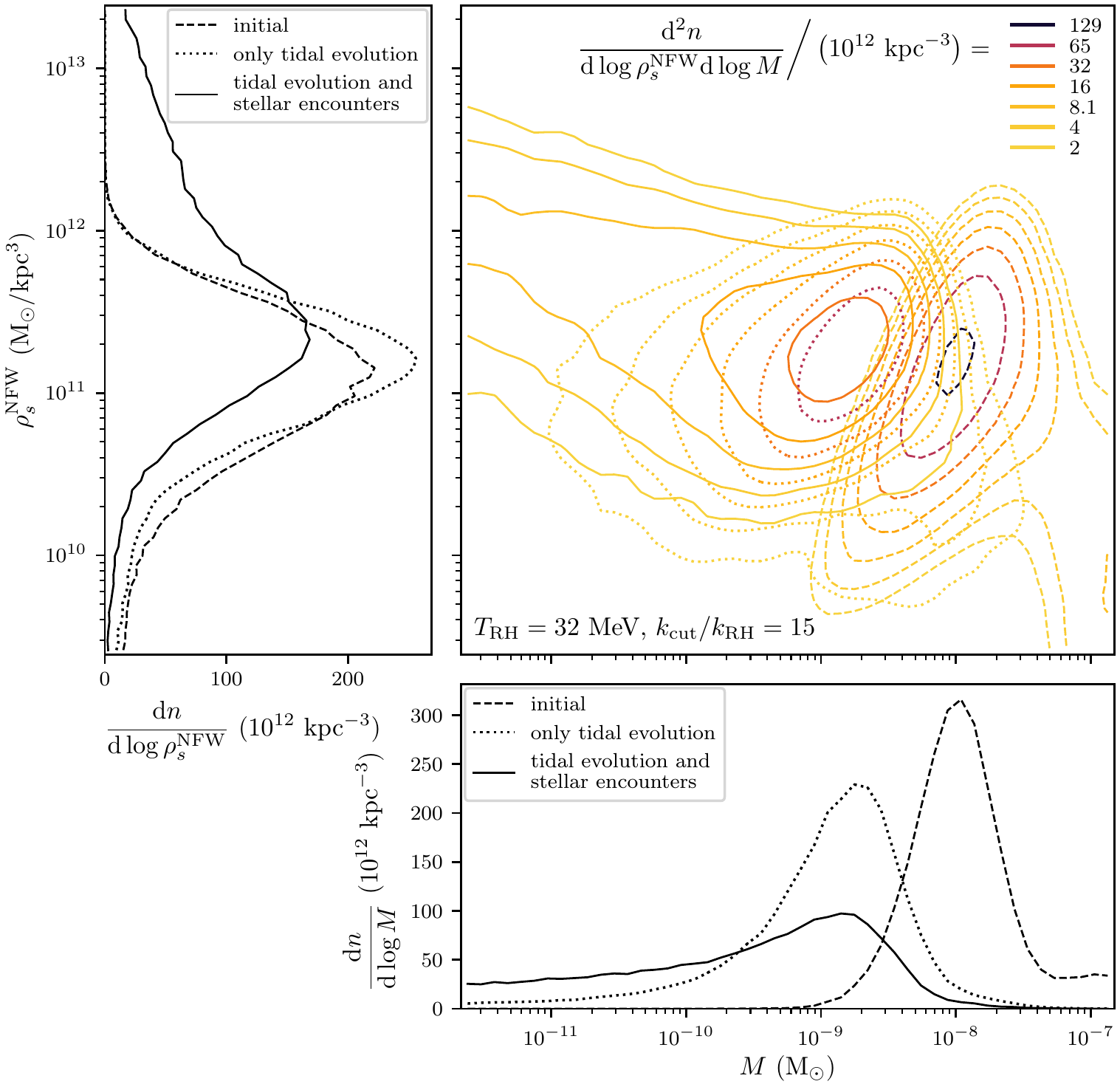}
	\caption{The distribution of microhalos near the Sun resulting from the EMD scenario with $\TRH=32$ MeV and $\xcut=15$. The main plot shows contours of the differential halo number density distributed in mass $M$ and NFW-equivalent scale radius $\rho_s^\mathrm{NFW}$ (see the text), while the left and bottom plots show the projected distributions in $\rho_s^\mathrm{NFW}$ and $M$ separately. In all plots, the dashed curves show the initial microhalo population before disruptive effects are accounted for (equivalently, before accretion onto a larger halo). The bulk of these halos are tightly clustered in $\rho_s^\mathrm{NFW}$--$M$ space because they form from the smallest-scale density variations above the free-streaming cutoff, but a tail of halos forming from larger-scale fluctuations is visible. Tidal forces from a host halo strip off the microhalos' weakly bound outskirts, resulting in halo populations with similar $\rho_s^\mathrm{NFW}$ but significantly reduced $M$ (dotted lines). Also including the impact of stellar encounters greatly spreads out the distribution (solid curves), producing a long tail of ``unlucky'' low-mass halos. Note that stellar encounters raise $\rho_s^\mathrm{NFW}$ only in the sense that stripping off a halo's low-density outskirts raises its average density. Stellar encounters still reduce a halo's density at any given radius.}
	\label{fig:halos15}
\end{figure*}

\begin{figure*}[tbp]
\centering
\includegraphics[width=.8\linewidth]{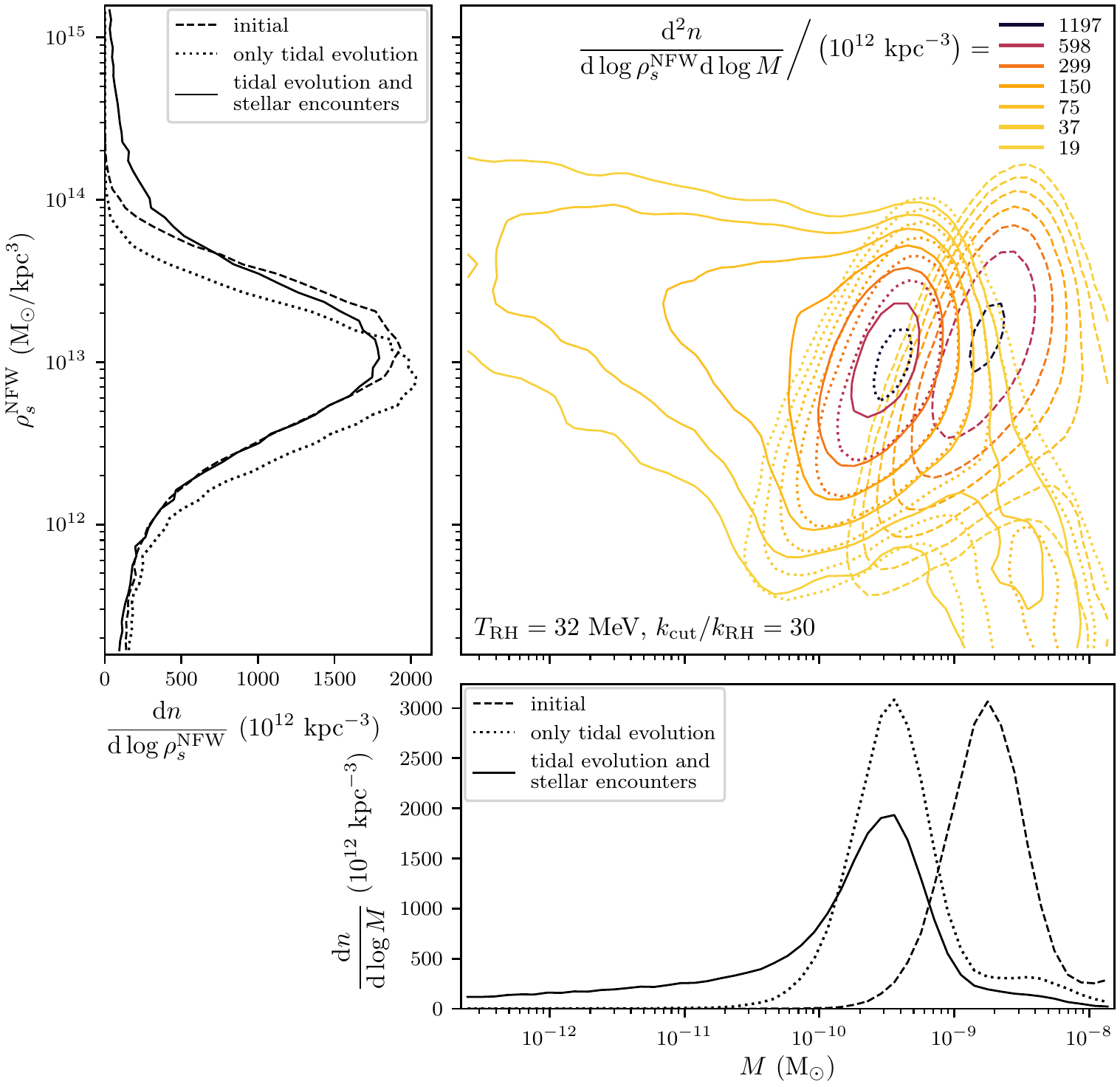}
\caption{Same as Fig.~\ref{fig:halos15} but assuming a smaller free-streaming scale. Here, $\xcut=30$. The initial halo population (dashed lines) exhibits coherently lower halo masses and higher internal densities but is otherwise the same as Fig.~\ref{fig:halos15}. However, the higher internal halo densities make this population less susceptible to disruptive tidal effects and stellar encounters (dotted and solid lines).}
\label{fig:halos30}
\end{figure*}

Figures \ref{fig:halos15} and~\ref{fig:halos30} show (as dashed lines) the initial microhalo distributions, computed as described above, for two different EMD scenarios. We show the distributions in NFW scale density $\rho_s^\mathrm{NFW}$ and halo mass $M$, where we take $M$ to be the virial mass at $z=2$, roughly the time at which microhalos might be expected to accrete onto the Galaxy.\footnote{For the initial halo population, we set $M=M_{200, m}$, the mass of the region that has $200$ times the background matter density. We compute $M$ solely for illustrative purposes; it has no impact on our later results.} The bulk of the microhalos form from density variations that are just above the free-streaming scale, and these halos comprise the main cluster in $\rho-M$ space. Beyond this, there is a tail of larger and less-dense halos that are associated with increasingly large-scale density variations.

\section{Disruption of microhalos}\label{sec:disruption}

In isolation, microhalos arising from EMD maintain static internal density profiles, only growing outward as they accrete material. In that case, the above predictions would remain accurate today. However, microhalos relevant to pulsar timing are in the vicinity of the Galactic disk at the present time. These microhalos accreted onto the Galactic halo and were subjected to evolution due to tidal forces and encounters with other objects. We now discuss how we treat these effects.

\subsection{Tidal evolution}

A common treatment for tidal evolution (e.g., Ref.~\cite{lee2021probing}) is to simply truncate a subhalo's density profile at its ``tidal radius'', the radius beyond which tidal forces from the host are stronger than the subhalo's self-gravity. While this procedure supplies a simple approximation, in reality tidal stripping is a continuous process as the subhalo revirializes in response to mass loss (e.g., Ref.~\cite{van2017disruption}). Instead, we use the model from Appendix E of Ref.~\cite{delos2019tidal}, which predicts the time evolution of $\rmax$ and $\mmax$ for an NFW subhalo orbiting an NFW host and is tuned to match idealized $N$-body simulations. The Galactic potential is not NFW, owing to the contribution of baryons, so as an approximation, we pick the NFW potential that best reproduces the observationally modeled Galactic rotation curves in Ref.~\cite{Bland_Hawthorn_2016} for Galactocentric radii smaller than $25$~kpc. This NFW potential has scale radius $r_s=3.4$~kpc and $\rho_s=4.0\times 10^8~\Msun/$kpc$^3$.\footnote{We emphasize that we employ the NFW profile solely to describe the Galactic potential and not its stellar distribution. We treat the Galaxy's stellar distribution more accurately later.}

This choice of host potential represents an approximation, and we note the following limitations. Galactic rotation curves are valid within the disk plane, but the dark matter halo---and hence microhalo orbits---typically extend far above and below this plane, where the potential is weaker. In this respect our choice is likely to overestimate the impact of tidal stripping. Meanwhile, abrupt features within the host potential heat subhalos, increasing the rate at which they lose mass. While we account for encounters with individual disk stars below, the encounter with the Galactic disk's bulk potential can itself be a significant factor in microhalo disruption. This concern suggests we may risk underestimating the impact of tidal stripping. However, Ref.~\cite{berezinsky2014small} estimates that the impact of individual stellar encounters dominates over that of coherent disk shocking for halos smaller than earth mass. As Fig.~\ref{fig:pk} indicates, EMD-induced microhalos are sub-Earth mass if $\TRH\gtrsim 10$~MeV. In any event, since our tidal evolution model is only calibrated for NFW host potentials, we leave to future work more precise modeling of microhalo survival within the Galactic potential.

We now sample the orbital parameters of microhalos at the solar radius $r_0=8$~kpc under the approximation that the microhalo velocity distribution is isotropic. Specifically, we use rejection methods to sample the orbital energy $E$ from the unnormalized distribution $\sqrt{E-\Phi(r_0)} f(E)$, where $f(E)$ is the phase-space distribution function and $\Phi$ is the Galactic potential. A fitting form for $f(E)$ is drawn from Ref.~\cite{widrow2000distribution}. The microhalo velocity's magnitude $|\vec V|$ immediately follows. To describe the direction of the microhalo's velocity $\vec V$, we uniformly sample $\mu\equiv\cos\theta$, where $\theta$ is the angle between $\vec V$ and the Galactic radius vector, and the angle $\phi$ between the microhalo's tangential velocity projection and the velocity $\vec V_\mathrm{LSR}$ of the local standard of rest. Since faster microhalos are encountered more frequently, we additionally use rejection methods to weight our orbit sample by
$V_\mathrm{rel}\equiv \left|\vec V-\vec V_\mathrm{LSR}\right|$.\footnote{\label{foot:pulsarv}In principle, we should weight the encounter frequency by the relative velocity with respect to each pulsar under consideration. However, the microhalo velocity dispersion is much larger than the dispersion in pulsar velocities, so pulsar motions are neglected in the analysis of timing distortions in Sec.~\ref{sec:procedure} and Ref.~\cite{lee2021probing}. In particular, the rms microhalo velocity with respect to the local standard of rest turns out to be about 330~km/s in our calculation, while the mean two-dimensional velocity of millisecond pulsars is measured to be about 87~km/s \cite{2005MNRAS.360..974H}.}

In this way, we associate a randomly sampled orbit with each initial microhalo from Sec.~\ref{sec:initial}. We now apply the tidal evolution model in Ref.~\cite{delos2019tidal} assuming a duration of 10~Gyr, which corresponds to accretion onto the Galaxy at a redshift of about $z=2$. This approximation does not have a major impact; large changes in the microhalo accretion redshift correspond to only modest changes in the tidal evolution duration.

\subsection{Stellar encounters}

The next step is to consider encounters with other objects, such as stars or other microhalos. Reference~\cite{delos2019breaking} found that the impact of encounters with other microhalos was subdominant to tidal evolution and stellar encounters even for the dark matter-dominated Draco dwarf spheroidal galaxy, so we can reasonably assume it is even more subdominant for microhalos in the solar neighborhood. However, stellar encounters are likely to have a significant, and perhaps even dominant, impact on microhalo evolution.

To handle stellar encounters, a common approximation (e.g., Ref.~\cite{schneider2010impact}) is to compare the energy injected into a microhalo by these encounters to the halo's total binding energy. However, as noted by Ref.~\cite{van2017disruption}, this computation is not clearly connected to the question of halo survival because these energy injections are inefficient: the most weakly bound particles receive the most energy in an encounter. Instead, we use the model in Ref.~\cite{delos2019evolution} to treat the impact of stellar encounters; this model precisely predicts the evolution of a halo's density profile due to impulsive point-particle encounters. To apply this model it is necessary that for each microhalo we sample a series of stellar encounters, and we do so as follows. Let $n_*(\vec R)$ be the number density of stars at the galactocentric position $\vec R$, and suppose a microhalo's orbital position and velocity are described by $\vec R(t)$ and $\vec V(t)$, respectively. The differential number of stellar encounters, per impact parameter $b$ and time $t$, is
\be\label{dNdt}
\frac{\diff^2 N_*}{\pi b\, \diff b\, \diff t} = n_*[\vec R(t)]\, \bar V_\mathrm{rel}[\vec V(t),\vec R(t)],
\ee
where
\be\label{Vrel}
\bar V_\mathrm{rel}(\vec V,\vec R) \equiv \int \diff^3 \vec V_* \left|\vec V_* - \vec V\right| f(\vec R; \vec V_*)
\ee
and $f(\vec R; \vec V_*)$ is the distribution of stellar velocities $\vec V_*$ at position $\vec R$. Appendix~\ref{sec:Vrel} shows how Eq.~(\ref{Vrel}) can be rapidly evaluated.

We use the Galactic model in Ref.~\cite{robin2003synthetic,*robin2004synthetic} to describe the distributions, $n_*(\vec R)$ and $f(\vec R; \vec V_*)$, of stellar positions and velocities, but for simplicity we cylindrically symmetrize the Galactic bar and assume that the stellar velocity dispersion is isotropic.\footnote{In principle, we should account for the time evolution of the Galaxy's stellar distribution. For instance, the bar should rotate and the stellar mass function should evolve. But for simplicity we approximate the Galaxy as static.}  With these distributions established, we use inverse transforms to sample stellar encounter times $t$ and impact parameters $b$ from Eq.~(\ref{dNdt}). We then sample the relative velocities $|\vec V_*-\vec V|$ of these encounters using the microhalo's orbital information and the stellar velocity dispersions. Finally, we sample the masses of the stars that our microhalos encounter from a Kroupa initial mass function with minimum mass $0.01~\Msun$ (we include brown dwarfs) and a high-mass index of 2.7 \cite{kroupa2002initial}.

The above considerations determine the stellar encounter distribution for each microhalo, and the next step is to apply the stellar encounter model of Ref.~\cite{delos2019evolution}. We additionally follow the suggestions of Ref.~\cite{delos2019breaking}, which carried out combined tidal evolution and stellar encounter simulations. To wit, we do not model any relaxation of microhalos between encounters, instead simply summing the (relative) energy injections from all encounters, and we apply the full sequence of stellar encounters after the tidal evolution. The tidal evolution model predicts for each halo its maximum circular velocity $v_\mathrm{max}$ and associated radius $r_\mathrm{max}$, and we relate these quantities back to a density profile by assuming that the profile after tidal evolution resembles the profile after stellar encounters, which, as discussed in Ref.~\cite{delos2019evolution}, is (almost) universal. We then apply the stellar encounter model with this density profile as a starting point. The post-encounter density profile given in Ref.~\cite{delos2019evolution} is $\rho=\rho_s (r_s/r) \exp\left[(-1/\alpha)\left(r/r_s\right)^\alpha\right]$ with $\alpha=0.78$, but as we discuss in Appendix~\ref{sec:profile}, this density profile becomes unphysically steep at large radii. We instead propose the form
\be\label{profile}
	\rho=\rho_s \frac{r_s}{r} \exp\!\left[-\frac{1}{\alpha}\!\left(\frac{r}{r_s}\right)^\alpha\!\! (1\!+\!q)^{1-\frac{1}{\beta}}\,_2F_1\!\left(\!1,1;1\!+\!\frac{1}{\beta};-q\!\right)\right]
\ee
with $q\equiv \left[(1/3)(r/r_s)^\alpha\right]^\beta$, $\alpha=0.78$, and $\beta=5$. Here, $_2F_1$ is the hypergeometric function. This profile transitions to $\rho\propto r^{-4}$ at large radii instead of becoming arbitrarily steep, and it agrees with the simulation we carry out in Appendix~\ref{sec:profile}.

Figures \ref{fig:halos15} and~\ref{fig:halos30} show the impact of tidal effects (dotted lines) and stellar encounters (solid lines) on the microhalo population for $\TRH=32$~MeV and two different free-streaming scales. The post-encounter density profile, Eq.~(\ref{profile}), has finite total mass, so we do not assume any radial truncation in determining each halo's mass. Additionally, to fairly compare characteristic density values between the disrupted and initial halo populations, we convert the post-encounter density profile's scale density $\rho_s$ into the equivalent NFW scale density $\rho_s^\mathrm{NFW}=\rho_s/1.17$ (see Ref.~\cite{delos2019evolution}). Microhalos evidently suffer a large drop in mass, which can be attributed to the loss of the halo's weakly bound outskirts. Much of this mass loss actually reflects the shift in density profile from NFW to Eq.~(\ref{profile}), but highly dense microhalos might not fully transition their density profiles, so we likely overestimate microhalo disruption when $\xcut$ is large.\footnote{In particular, in the $\xcut=30$ case the tidal-stripping model predicts essentially no evolution in the scale parameters of the microhalo density profiles, so almost all of the change between the dashed and dotted lines in Fig.~\ref{fig:halos30} arises from the change in the density profile's form. Thus, it is likely that for $\xcut\gtrsim 30$ the assumption that density profiles fully transition into the new form represents an overestimate. However, it should be noted that the total microhalo mass $M$ does not enter into the pulsar-timing analysis in Sec.~\ref{sec:procedure} except for halos that remain very distant from the pulsar. The pulsar-timing signal is dominated by the closest encounters \cite{lee2021probing}, for which only the halo mass interior to the pulsar's separation from the halo is relevant.} Meanwhile, the characteristic density $\rho_s^\mathrm{NFW}$ of microhalos increases because when the low-density outskirts are stripped, the average density of the halo rises.\footnote{As can be seen in Ref.~\cite{delos2019evolution}, the density at any given radius within a halo always drops due to stellar encounters even as the halo's characteristic density rises.} This change is relatively modest and arises predominantly due to stellar encounters and not tidal stripping. The stochastic nature of stellar encounters also spreads out the microhalo distribution, producing a long tail of low-mass halos that suffered more disruption by chance.

\section{Pulsar timing signals}\label{sec:procedure}

With microhalo populations established, the remaining step is to determine the extent to which they can be detected by PTAs. In particular, we consider here the timing distortions that arise from the Doppler effect when microhalos perturb pulsar motions. While there is also a Doppler effect associated with perturbations to the Earth's motion, Ref.~\cite{lee2021probing} found that this effect generally produces weaker constraints (although it has the potential to probe smaller masses). Timing distortions could also arise from the Shapiro effect when microhalos cross the line of sight to a pulsar, but this effect is most sensitive to mass scales $M\gtrsim 10^{-5}~\Msun$ \cite{Ramani_2020}, which are larger than the masses of EMD-induced microhalos (see Fig.~\ref{fig:pk}).

We follow the procedure of Ref.~\cite{lee2021probing} and use a modified version of the associated Monte Carlo simulation code.\footnote{The original code resides at the URL \url{https://github.com/szehiml/dm-pta-mc}, and our modified code is located at \url{https://github.com/delos/dm-pta-mc}.} This code prepares a random array of pulsar positions, and for each pulsar, it randomly samples encounters with microhalos. Pulsar velocities (relative to the local standard of rest) are neglected because they are much smaller than microhalo velocities (see footnote~\ref{foot:pulsarv}). We modified the code to sample microhalos from the distribution computed in Secs. \ref{sec:initial} and~\ref{sec:disruption}. The remainder of this section constitutes a review of the computational procedure described in Ref.~\cite{lee2021probing} through which we evaluate the the pulsar timing signal that arises from such a microhalo distribution.

A pulsar's Doppler shift due to microhalo encounters manifests into an accumulated shift $\delta \phi$ in the pulse phase. For a pulsar with frequency $\nu$ and earth-pulsar direction vector $\uvec d$, the phase shift due to an encounter with a point object (like a primordial black hole) with mass $M$, velocity $\vec v$, and impact parameter $\vec b$, is
\be\label{phaseshift_pt}
\delta \phi_\mathrm{pt}
=
\frac{G M \nu}{v^{2}c} \uvec d \cdot\left(\sqrt{1+x^{2}} \uvec b -\sinh ^{-1}(x) \uvec v \right)
\ee
(where $c$ is the speed of light). Here, $x \equiv (b/v)(t-\tilde t)$, where $\tilde t$ is the time of closest approach when the vector from the pulsar to the object is $\vec b$. The phase shift due to an extended object must in general be computed numerically, but to avoid the computational expense, we follow Ref.~\cite{lee2021probing} and conservatively approximate the phase shift due to a microhalo encounter as
\be\label{phaseshift_ext}
\delta \phi
\simeq
\delta \phi_\mathrm{pt} \mathcal{F}(r_\mathrm{min}).
\ee
Here, $r_\mathrm{min}\geq b$ is the microhalo's closest approach to the pulsar during the observing time, and the ``form factor''
\be
\mathcal{F}(r)\equiv M(r)/M\leq 1,
\ee
where $M(r)$ is the enclosed mass profile, is the factor by which the microhalo's finite extent scales its gravitational field at the distance $r$, relative to a point object.

\begin{figure}[tbp]
	\centering
	\includegraphics[width=\columnwidth]{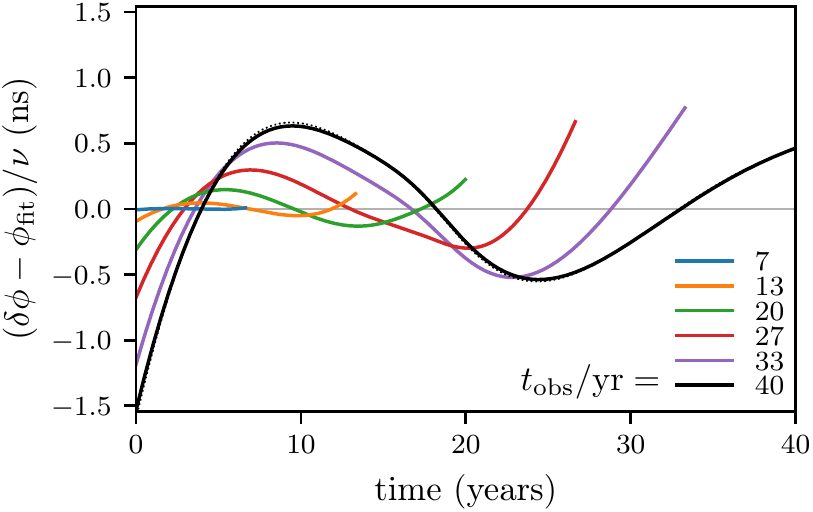}
	\caption{An example pulsar phase-shift signal $\delta\phi-\delta\phi_\mathrm{fit}$, as a function of time, due to microhalos arising from an EMD scenario with $\TRH=32$~MeV and $\xcut=30$. A quadratic fit has been subtracted as described in the text. Different curves show the signals arising from \textit{the same set of microhalo encounters} for a range of observation durations $\tobs$. These curves do not overlap only because the subtracted quadratic fit is different in each case. For the 40-year curve (black) we also show, as a thin dotted line, the phase shift computed exactly instead of using the approximation in Eq.~(\ref{phaseshift_ext}). This curve is barely visible due to how little it differs from the approximate result (solid line), which suggests that the approximation does not have a major impact.}
	\label{fig:signal}
\end{figure}

We sum the microhalo-induced phase shift over all of the randomly sampled microhalos in the pulsar's neighborhood. Next, we subtract a quadratic fit, $\phi_\mathrm{fit}(t)=\phi_{0}+\phi_{1} t+\phi_{2} t^{2}$, reflecting that these low-order terms are degenerate with the pulsar's intrinsic behavior.\footnote{In particular, we follow Refs.~\cite{Dror_2019,Ramani_2020,lee2021probing} in assuming that $\ddot\nu$ and higher derivatives of the pulsar frequency arise solely due to perturbations by substructure. In practice contributions to $\ddot\nu$ can arise from the pulsar's intrinsic spin-down, its unperturbed line-of-sight velocity, or a combination of the two; these contributions can be estimated and are of order $\ddot\nu/\nu\sim 10^{-31}$~s$^{-2}$ \cite{2018MNRAS.478.2359L}.} We plot in Fig.~\ref{fig:signal} an example phase-shift signal from a single pulsar due to a microhalo distribution arising from the EMD scenario with $\TRH=32$~MeV and $\xcut=30$. A range of observation durations $\tobs$ are shown (for the same microhalo encounters); the signal is different in each case due to the subtraction of the quadratic fit. While these phase shifts $\delta\phi$ are evaluated using the approximate expression in Eq.~(\ref{phaseshift_ext}), we also compare the exact phase shift computed using numerical integration (dotted black curve, nearly overlapping the solid black curve). The difference is marginal, which suggests that the approximation does not have a large impact.

We repeat the above calculation for all $N_P$ pulsars under observation. Additionally, to account for stochasticity, we repeat this process for 1000 ``universes'' each with independently sampled microhalo distributions. From each pulsar's phase-shift signal, we compute the signal-to-noise ratio (SNR) as described in Ref.~\cite{lee2021probing} assuming optimal filtering, and the SNR associated with each ``universe'' is taken to be the largest SNR associated with any pulsar therein. The overall SNR is then the tenth percentile of the SNRs of the 1000 ``universes''. Finally, we evaluate the significance level $\sigma$ associated with that SNR. That is, $\sigma$ is the significance level with which we can exclude the possibility that the given SNR arose from noise alone. These choices are identical to those made in Ref.~\cite{lee2021probing}.

\section{Results and discussion}\label{sec:results}

We initially fix the observational cadence to 1~week and the rms timing residual to 10~ns, and we assume that the timing noise is white. These are the choices made in Ref.~\cite{lee2021probing}, but we discuss their impact and validity later. However, we vary the observing time $\tobs$ and the number of pulsars $N_P$ under observation, and we also vary the cosmology parameters $\TRH$ and $\xcut$. We thus have a four-dimensional parameter space, and each point in this space is associated with a particular pulsar timing SNR and hence statistical significance, as described in Sec.~\ref{sec:procedure}. From a more practical point of view, for any given pulsar count $N_P$ and observation duration $\tobs$, we may target a minimum significance level for a pulsar timing signal. There is a two-dimensional subspace of cosmological parameters $\TRH$ and $\xcut$ that satisfy this constraint and hence can be probed by the given observational parameters.

\subsection{Impact of pulsar count and observation time}

\begin{figure}[tbp]
\centering
\includegraphics[width=\columnwidth]{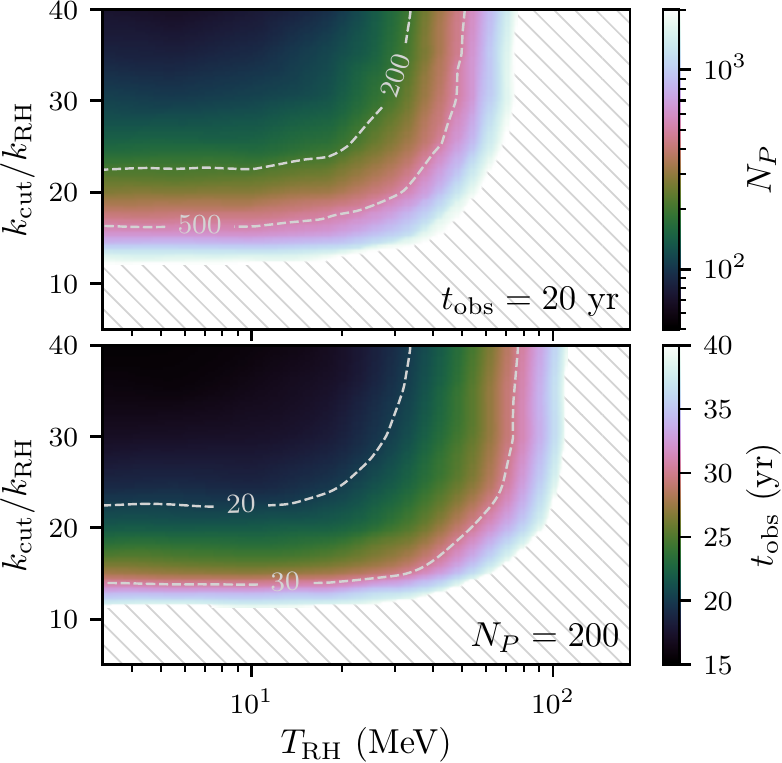}
\caption{
\textit{Top}: The pulsar count $N_P$ required to detect microhalos arising from EMD scenarios with each $\TRH$ and $\xcut$ at 2$\sigma$ significance, if the observation duration is fixed at $\tobs=20$~years. In this case at least 70 pulsars are needed to probe any EMD scenario with $\TRH>3$~MeV and $\xcut<40$.
\textit{Bottom}: Similarly, the observation duration $\tobs$ required to probe each EMD scenario if we instead fix $N_P=200$. In this case at least 15~years of observation are required.
In both cases, more pulsars or observation time allow higher $\TRH$ and smaller $\xcut$ to be probed. The hatched region requires $N_P>2000$ (top) or $\tobs>40$~years (bottom).}
\label{fig:NPtobs}
\end{figure}

We first explore the impact of the observation duration $\tobs$ and pulsar count $N_P$ on the range of EMD scenarios that can be probed. We begin by varying $N_P$ with fixed $\tobs=20$~years. The upper panel of Fig.~\ref{fig:NPtobs} shows, as colors, the number of pulsars $N_P$ needed to detect microhalos arising from the given $\TRH$ and $\xcut$ at $2\sigma$ significance. With this $\tobs$, the detection of microhalos arising from the EMD scenario with $\TRH=3$~MeV and $\xcut=40$ requires at least $N_P\simeq 70$~pulsars, a number comparable to the count in existing PTAs (e.g., Ref.~\cite{2019MNRAS.490.4666P}). Meanwhile, as $N_P$ is raised, higher reheat temperatures $\TRH$ and lower cutoff ratios $\xcut$ can be probed. This trend is easy to understand: higher $\TRH$ leads to microhalos of lower mass, while lower $\xcut$ leads to microhalos of lower internal density; both of these trends make the microhalos more difficult to detect. Scenarios with $\TRH\gtrsim 70$~MeV or $\xcut\lesssim 13$ require more than 2000 pulsars if $\tobs=20$~years.

We next explore what observation duration $\tobs$ is necessary if we are able to observe $N_P=200$ pulsars. The lower panel of Fig.~\ref{fig:NPtobs} shows, again as colors, the $\tobs$ needed to detect microhalos arising from the EMD scenario with the given $\TRH$ and $\xcut$ at $2\sigma$ significance. With this pulsar count, at least $\tobs\simeq 15$~years (black) are required to detect microhalos arising from the EMD scenario with $\TRH=3$~MeV and $\xcut=40$. Similarly to $N_P$, raising $\tobs$ allows higher $\TRH$ and lower $\xcut$ to be probed. EMD scenarios with $\xcut\lesssim 10$ or $\TRH\gtrsim 100$~MeV require more than 40~years of observation time if $N_P=200$.

Na\"ively one might expect $N_P$ and $\tobs$ to have a similar quantitative impact, because the effective spatial volume that pulsar timing probes is proportional to both. That is, the number of microhalos passing near any given pulsar is proportional to $\tobs$, so the total number of microhalos passing near pulsars is proportional $N_P\tobs$. However, by comparing the two panels of Fig.~\ref{fig:NPtobs} we notice that raising the observation duration $\tobs$ has a much greater impact than raising the pulsar count $N_P$ on the range of EMD scenarios that can be probed. For instance, raising $\tobs$ from 20~years to 30~years has about as much impact as boosting $N_P$ by a factor of 10 from 200 to 2000.

The reason for observation time's power is that beyond boosting the number of microhalo encounters, $\tobs$ has another advantage that is illustrated in Fig.~\ref{fig:signal}. This figure suggests that raising $\tobs$ increases not only the duration of the signal but also its amplitude. The mathematical reason for this effect is the subtraction of the quadratic fit. Physically, more observation time allows us to better distinguish the impact of microhalos---which induce perturbations to the phase shift $\delta\phi$ at higher-than-quadratic order---from the pulsar's intrinsic quadratic-order phase shift.\footnote{In practice, long observation durations can also increase the impact of timing noise if the timing residuals are correlated over long time periods. For instance, the contribution from any intrinsic frequency second derivative, $\ddot\nu$, could lead to such correlations. The analysis here does not reflect this possibility because we follow Ref.~\cite{lee2021probing} in assuming, for simplicity, that the pulsars' intrinsic timing noise is white.} Put another way, the time scale associated with microhalo encounters is of order years (see Fig.~\ref{fig:signal}), so tens of years of observation time are needed to fully resolve them.

\subsection{Detection prospects for EMD reheat temperatures and cutoff ratios}

Figure~\ref{fig:NPtobs} indicates that when $\xcut\gtrsim 30$, the detection prospects of an EMD scenario are nearly independent of $\xcut$, and when when $\TRH\lesssim 20$~MeV, the detection prospects are nearly independent of $\TRH$. Both of these trends can be understood in light of Fig.~2 of Ref.~\cite{lee2021probing}. Low $\TRH$ corresponds to large microhalo mass scales, and this figure indicates that when the mass scale of microhalos exceeds roughly $10^{-7}~\Msun$, their detection prospects become mass independent. Meanwhile, high $\xcut$ leads to microhalos of high internal density, which means that these halos are highly centrally concentrated. In this case we approach the point-mass limit, in which microhalo detection prospects become independent of their level of central concentration.

\begin{figure}[tbp]
\centering
\includegraphics[width=\columnwidth]{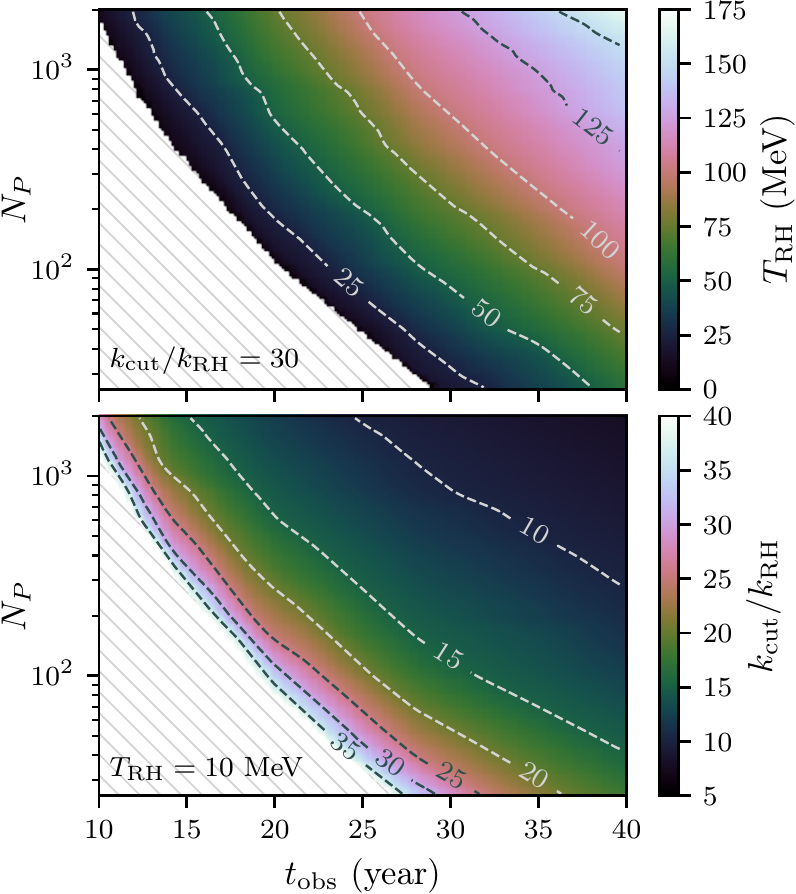}
\caption{The maximum reheat temperature $\TRH$ (top) or the minimum cutoff ratio $\xcut$ (bottom) for which the resulting microhalo population is detectable at $2\sigma$ significance with the given pulsar count $N_P$ and observation duration $\tobs$. In the upper panel we fix $\xcut=30$, but the result applies broadly to the case $\xcut\gtrsim 30$ since microhalo detection prospects are fairly insensitive to $\xcut$ in this regime. In the lower panel we fix $\TRH=10$~MeV, but the result applies generally to the case $\TRH\lesssim 20$~MeV since detection prospects are insensitive to $\TRH$ in this regime.}
\label{fig:TRHxcut}
\end{figure}

Thus, in these regimes we can express EMD detection prospects as a function of $\TRH$ alone and $\xcut$ alone, respectively. Accordingly, the upper panel of Fig.~\ref{fig:TRHxcut} shows the number $N_P$ of pulsars and observation duration $\tobs$ needed to detect microhalos (at $2\sigma$ significance) arising from an EMD scenario with a given $\TRH$ in the case where $\xcut\gtrsim 30$.\footnote{For $\xcut\gtrsim 40$, microhalos can form significantly before the onset of the last matter-dominated epoch. Halos can form during radiation domination and would be even denser than those that form later \cite{2019PhRvD.100j3010B}, so our conclusions still apply in this case. However, if $\xcut$ is sufficiently large that halos form during the EMD epoch, their evaporation at reheating can suppress small-scale structure \cite{2019PhRvD.100j3010B,2021arXiv210710293B}. Further analysis is needed to determine microhalo detection prospects in these cosmologies.} Evidently, microhalos arising from reheat temperatures as high as $\TRH\simeq 170$~MeV can be detected if $N_P=2000$ and $\tobs=40$~years. We also remark that to the extent that a direct comparison can be made, our results are similar to the EMD detection prospects presented in Ref.~\cite{lee2021probing}.\footnote{For instance, in Fig.~8 of Ref.~\cite{lee2021probing} the observation duration is fixed at $\tobs=30$~years, and when $N_P=1000$, $2\sigma$ significance is achieved when $\TRH$ is slightly over 100~MeV, a result that is also evident in our Fig.~\ref{fig:TRHxcut}.}

Similarly, the lower panel of Fig.~\ref{fig:TRHxcut} shows the $N_P$ and $\tobs$ required to detect microhalos (at $2\sigma$ significance) arising from an EMD scenario with a given $\xcut$ if $\TRH\lesssim 20$~MeV. In this case, microhalos associated with cutoff ratios as low as $\xcut\simeq 8$ can be detected if $N_P=2000$ and $\tobs=40$~years.

\subsection{Impact of timing noise and cadence}\label{sec:noise}

\begin{figure}[tbp]
\centering
\includegraphics[width=\columnwidth]{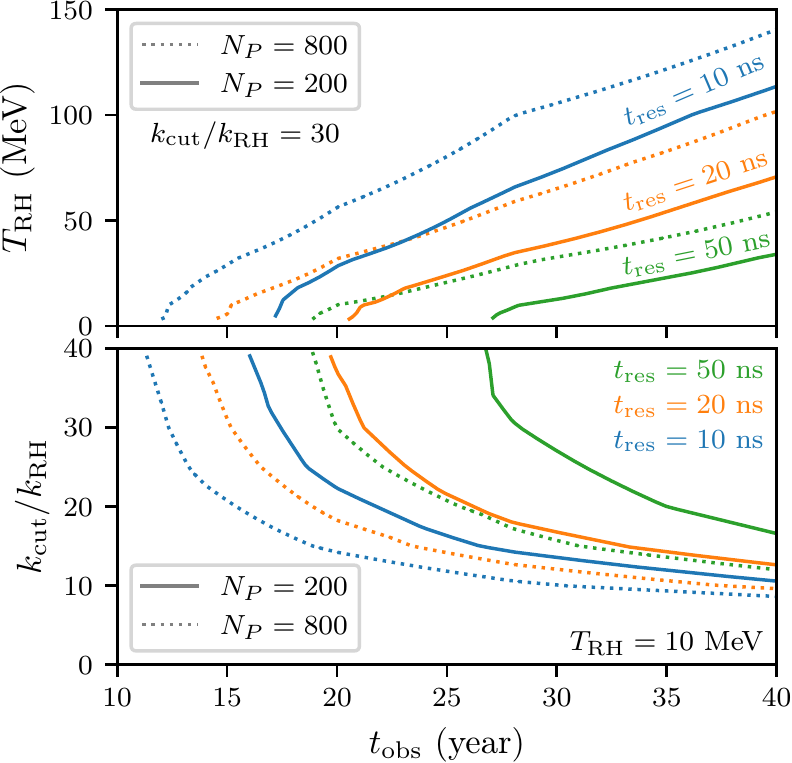}
\caption{Impact of the rms timing residual $\tres$ on the maximum reheat temperature $\TRH$ (top) or the minimum cutoff ratio $\xcut$ (bottom) that can be probed for the given $N_P$ and $\tobs$. Specifically, these curves are the slices of Fig.~\ref{fig:TRHxcut} at fixed $N_P=200$ (solid curves) and $N_P=800$ (dashed curves), but we consider several different values of $\tres$: 10~ns (blue), 20~ns (orange), and 50~ns (green).}
\label{fig:noise}
\end{figure}

Up to this point, we have fixed the observational cadence at \mbox{$\tcad=1$~week} and the rms timing noise residual at \mbox{$\tres=10$~ns}. This is an optimistic timing-noise assumption; timing noise residuals in present PTAs are on the order of $\mu$s or greater (e.g., Ref.~\cite{2019MNRAS.490.4666P}), although precision modeling can mitigate this noise (e.g., Ref.~\cite{2021MNRAS.502..478G}). In Fig.~\ref{fig:noise} we test how EMD detection prospects change if $\tres$ is increased. This figure shows, for $\tres$ ranging from 10 to 50~ns, the maximum $\TRH$ and minimum $\xcut$ detectable for given $\tobs$ and $N_P$. Evidently, the rms timing residual $\tres$ is a highly important variable, and $\tres\lesssim 0.1$~$\mu$s is necessary to detect microhalos arising from EMD using pulsar timing if $\tobs\lesssim 40$~years and $N_P\lesssim 800$.

Under the assumptions made in Sec.~\ref{sec:procedure} (and Ref.~\cite{lee2021probing}), decreasing the observational cadence $\tcad$ can counteract the impact of timing noise because the signal-to-noise ratio scales as $\mathrm{SNR}\propto \tres^{-1}\tcad^{-1/2}$ (see Eq.~14 of Ref.~\cite{lee2021probing}). That is, since the noise is assumed to be white (uncorrelated), its impact can be reduced to an arbitrary extent by simply taking more measurements. In practice, however, a significant contribution to the timing noise is red noise associated with the pulsar's spin (e.g., Refs.~\cite{2010ApJ...725.1607S,2015JPhCS.610a2019W}), which exhibits temporal correlations. For a red noise spectrum there is less advantage to raising the observational cadence.

We also remark that the timescale associated with a typical microhalo encounter is about a year if $\TRH\sim 200$~MeV and is longer for lower $\TRH$. Any cadence $\tcad$ much shorter than a year is sufficient to temporally resolve these encounters. Consequently, the cadence plays no other role in our calculation than to control the noise level per the above discussion.

\section{Conclusion}\label{sec:conclusion}

In this article, we explored the prospects for pulsar timing arrays to probe early matter-dominated eras through their impact on small-scale dark structure. We found that in order to detect EMD-induced dark matter microhalos, pulsar timing noise residuals must be brought below roughly the $0.1$~$\mu$s level. Additionally, due to the long timescale associated with microhalo-pulsar encounters (of order years), tens of years of observing time are necessary. However, with 10~ns timing residuals, EMD scenarios can begin to be probed with about 20~years of observation time and as few as 70~pulsars. This is roughly the pulsar count in existing arrays \cite{2019MNRAS.490.4666P}. With 40~years of observing time and hundreds to thousands of pulsars, EMD reheat temperatures $\TRH$ up to approximately 150~MeV can be reached. Note that detection prospects are only weakly sensitive to the pulsar count.

Figure~\ref{fig:TRHxcut} shows our main results. In addition to the reheat temperature $\TRH$, which sets the mass scale for EMD-induced microhalos, we also explore the detection prospects with respect to the dark matter's free streaming scale $\kcut$. Specifically we consider the ratio $\xcut$, where $\kRH$ is the wavenumber that enters the horizon at reheating. The combination $\xcut$ is fundamentally important to the microhalo population induced by early matter domination because it sets microhalos' internal density scale (see Figs.~\ref{fig:pk}--\ref{fig:halos30}). For $\xcut\gtrsim 30$ microhalos are close to the point-mass limit, for the purpose of pulsar timing distortions caused by the Doppler effect. In this case the detectable range of $\TRH$ is insensitive to $\xcut$. For smaller $\xcut$ microhalos' spatial extent matters, and $\xcut$ as small as about $8$ can be probed given again hundreds to thousands of pulsars and 40~years of observing time (see Fig.~\ref{fig:TRHxcut}).

This work refines the EMD detection prospects presented in Ref.~\cite{lee2021probing}. The major new feature of our analysis is the use of the recently developed semianalytic microhalo models put forth in Refs.~\cite{delos2019predicting,delos2019tidal,delos2019evolution}. These models describe the formation and evolution of the first and smallest dark matter halos, and they were specifically built for the purpose of understanding the microhalo populations that arise from EMD scenarios and other cosmologies that feature boosted small-scale inhomogeneity. The application of these models is discussed in Secs. \ref{sec:initial} and~\ref{sec:disruption}, and \textsc{Python} codes that implement them are publicly available.\footnote{\url{https://github.com/delos/microhalo-models}.} Compared to Ref.~\cite{lee2021probing}, we also explore more extensively the space of cosmological and observational parameters. Consequently, we are able to clarify EMD detection prospects more precisely and to do so in terms of both $\TRH$ and $\xcut$.

We close with more general remarks about the capacity for PTAs to probe the Universe's early history. The results of a simplified analysis in Ref.~\cite{lee2021probing} suggest that pulsar timing is sensitive to halo masses above approximately $2\times 10^{-8}~\Msun$ (see Fig. 2 of that article). Through the size of the cosmological horizon, this mass scale corresponds to a time when the temperature of the Universe was roughly 150~MeV. Consequently, one can make a general statement that PTAs are able to probe the early Universe up to a temperature of about 150~MeV. EMD is not the only early-Universe scenario that results in boosted small-scale power. For instance, pulsar timing could also probe the possibility of domination by a \textit{cannibal} species with number-changing interactions, which leaves a different spectral imprint \cite{2021PhRvD.103j3508E,2021arXiv210609041E}, as long as it occurred below 150~MeV. If the dark matter is an axion and its Peccei-Quinn phase transition occurred below 150~MeV, then the resulting \textit{axion miniclusters} \cite{1988PhLB..205..228H,1994PhRvD..50..769K,1996ApJ...460L..25K,2017PhRvL.119b1101F,2018PhRvD..97h3502F,2020AJ....159...49D,2020arXiv201105377K,2021PhRvD.104b3515X} could also be detected using pulsar timing. We anticipate that the procedure through which we characterized the microhalo population arising from EMD---and hence the resulting pulsar timing signals---could be straightforwardly applied to these other scenarios as well.

\section*{Acknowledgements}
We thank Simon White for useful discussions. TL is partially supported by the Swedish Research Council under contract
2019-05135, the Swedish National Space Agency under
contract 117/19 and the European Research Council under grant 742104.

\appendix

\section{Integrating the stellar encounter velocity distribution}
\label{sec:Vrel}

In this appendix we evaluate the integral in Eq.~(\ref{Vrel}) that expresses the average relative velocity $\bar V_\mathrm{rel}(\vec V,\vec R)$ between a microhalo at position $\vec R$ and velocity $\vec V$ and the stars it encounters. Since we assume a spherically symmetric potential, a microhalo's orbit is planar with a static angle $\phi$ to the Galactic disk plane. The microhalo's position and velocity within this plane are described by the radius $R$, angle $\theta$, and associated velocity components $V_r$ and $V_\theta$, all of which are functions of time. We can decompose the halo's velocity instead into components
\be
V_{\parallel} = \frac{\cos\phi}{\cos\theta}\left(1+\tan^2\theta\cos^2\phi\right)^{-1/2}V_\theta
\ee
parallel to the mean stellar motion, which is assumed to be circular within the Galactic plane, and 
\be
V_{\perp} = \left(V_r^2+\frac{\sin^2\phi}{1+\tan^2\theta\cos^2\phi}V_\theta^2\right)^{1/2}
\ee
perpendicular thereto. If the stars have mean velocity $\bar V_*$ and isotropic velocity dispersion $\sigma$ (both of which can be functions of position), then Eq.~(\ref{Vrel}) can be written
\be
\bar V_\mathrm{rel}(\vec V,\vec R) = \sigma\, F\!\left(\sigma^{-1}\sqrt{\left(V_\parallel-\bar V_*\right)^2+V_\perp^2}\right).
\ee
Here, 
\be
F(a)\equiv \int\!\! \frac{\diff^3 \vec x}{(2\pi)^{3/2}} \sqrt{(x-a)^2+y^2+z^2}\, \e^{-|\vec x|^2/2}
\ee
with $\vec x\equiv (x,y,z)$, which can be rapidly evaluated using an interpolation table. We also note that $F(a)\simeq a+1/a$ to within one part in $10^4$ when $a>3$.

\section{The impulsively stripped density profile}
\label{sec:profile}

\begin{figure}[tbp]
\centering
\includegraphics[width=\columnwidth]{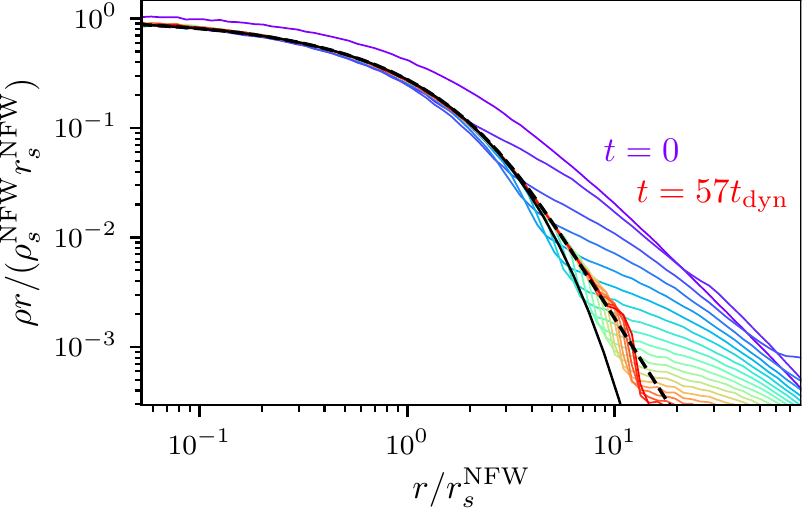}
\caption{The density profile of a simulated microhalo (initially NFW with parameters $\rho_s^\mathrm{NFW}$ and $r_s^\mathrm{NFW}$) as it responds to an impulsive stellar encounter at time $t=0$. Colored curves show the profile in time intervals of $3t_\mathrm{dyn}$, with $t_\mathrm{dyn}$ defined as in Ref.~\cite{delos2019evolution}. The solid black line shows Eq.~(\ref{profile_base}), the expression proposed in Ref.~\cite{delos2019evolution} as a universal fit to the post-encounter density profile. This expression is fitted (with $\rho_s$ and $r_s$ allowed to vary) to the late-time simulated density profile up to $r=4r_s^\mathrm{NFW}$. The dashed line shows the modified profile, Eq.~(\ref{profile_}), with the same $\rho_s$ and $r_s$ (not an independent fit); it is identical to Eq.~(\ref{profile_base}) at small radii but approaches $\rho\propto r^{-4}$ at large radii, as it should \cite{jaffe1987envelopes}. Moreover, tuned such that $\beta=5$, it also accurately matches the temporally converged part of the simulated density profile at all radii.}
\label{fig:profile}
\end{figure}

Based on simulations of microhalos undergoing impulsive stellar encounters, Ref.~\cite{delos2019evolution} proposed
\begin{equation}\label{profile_base}
	\rho=\rho_s \frac{r_s}{r} \exp\left[-\frac{1}{\alpha}\left(\frac{r}{r_s}\right)^\alpha\right],
\end{equation}
with $\alpha=0.78$, as the universal density profile of a halo initially possessing an NFW profile after arbitrarily many encounters. However, this profile becomes arbitrarily steep at large radii, whereas an analytic argument by Ref.~\cite{jaffe1987envelopes} suggests that the post-encounter profile does not steepen beyond $\rho\propto r^{-4}$. The simulations in Ref.~\cite{delos2019evolution} each covered a time duration equal to only about 10 times the dynamical timescale within the halo's scale radius, and within that duration the large-radius density profile did not stabilize sufficiently to test this argument. Consequently, in this appendix we carry out a new impulsive encounter simulation using the methodology of Ref.~\cite{delos2019evolution}. The simulated microhalo initially has an NFW density profile with scale parameters $r_{s,0}$ and $\rho_{s,0}$, and it undergoes an encounter with a star of mass $M_*$ at impact parameter $b$ and relative velocity $V$ such that $M_*/(V b^2) = 0.52\sqrt{\rho_{s,0}/G}$ and $b=16r_{s,0}$. This description completely characterizes an encounter in the limit that the encounter is impulsive. We follow the halo's response for almost 60 dynamical time intervals.

Figure~\ref{fig:profile} shows how the microhalo's density profile evolves in this simulation. To fit the time-stabilized part of this density profile at late times, we propose the form
\begin{equation}\label{profile_}
\rho=\rho_s \frac{r_s}{r} \exp\!\left[-\frac{1}{\alpha}\!\left(\frac{r}{r_s}\right)^\alpha\!\! (1\!+\!q)^{1-\frac{1}{\beta}}\,_2F_1\!\left(\!1,1;1\!+\!\frac{1}{\beta};-q\!\right)\right]
\end{equation}
with $q\equiv \left[(1/3)(r/r_s)^\alpha\right]^\beta$, $\alpha=0.78$, and $\beta=5$.\footnote{Equation~(\ref{profile_}) is obtained by integrating the expression $\diff\ln\rho/\diff\ln r = -1 - \left[(r/r_s)^{-\alpha\beta} + 3^{-\beta}\right]^{-1/\beta}$, which interpolates smoothly from Eq.~(\ref{profile_base}) at small radii to $\rho\propto r^{-4}$ at large radii.} Here, $_2F_1$ is the hypergeometric function. This profile is identical to Eq.~(\ref{profile_base}) at small radii but transitions to $\rho\propto r^{-4}$ at large radii. The suddenness of the transition is controlled by the parameter $\beta$, and we fix it through comparison with our simulation. As shown in Fig.~\ref{fig:profile}, Eq.~(\ref{profile_base}) (solid black curve) fails to match the post-encounter density profile at large radii because the latter does not steepen beyond $\rho\propto r^{-4}$ at large radii. Instead, Eq.~(\ref{profile_}) (dashed curve) accurately describes the post-encounter density profile.

\bibliography{main}
\end{document}